\documentclass{ieeetran}

\usepackage{microtype}
\usepackage{graphicx}
\usepackage{subfigure}
\usepackage{booktabs}

\usepackage{times}
\usepackage{epsfig}
\usepackage{graphicx}
\usepackage{amsmath}
\usepackage{amssymb}
\usepackage{enumitem}
\usepackage{hhline}
\usepackage{soul}
\usepackage{import}

\usepackage[pagebackref=true,breaklinks=true,colorlinks,bookmarks=false]{hyperref}

\usepackage{color}
\usepackage{multirow}
\usepackage[normalem]{ulem}
\usepackage{amsmath}

\usepackage{xspace}

\newcommand{\sikha}[1]{\textcolor{blue}{#1}}

\def\baselinestretch{0.98}

\DeclareRobustCommand{\hlgreen}[1]{{\sethlcolor{green}\hl{#1}}}

\newcommand{\adv}{\ensuremath{\mathcal{A}}\xspace}

\newcommand{\pmul}{\ensuremath{\pi_{\mathsf{DM}}}\xspace}
\newcommand{\pmmul}{\ensuremath{\pi_{\mathsf{DMM}}}\xspace}

\newcommand{\pargmax}{\ensuremath{\pi_{\mathsf{ARGMAX}}}\xspace}
\newcommand{\prelu}{\ensuremath{\pi_{\mathsf{RELU}}}\xspace}

\newcommand{\plt}{\ensuremath{\pi_{\mathsf{LT}}}\xspace}

\newcommand{\pdiv}{\ensuremath{\pi_{\mathsf{DIV}}}\xspace}
\newcommand{\pfselect}{\ensuremath{\pi_{\mathsf{FSELECT}}}\xspace}
\newcommand{\pvc}{\ensuremath{\pi_{\mathsf{LABELVIDEO}}}\xspace}

\newcommand{\pisoft}{\ensuremath{\pi_{\mathsf{SOFT}}}\xspace}

\newcommand{\pifi}{\ensuremath{\pi_{\mathsf{FINFER}}}\xspace}

\newcommand{\PPFRAMESELECT}{\ensuremath{\pi_{\mathsf{FSELECT}}}\xspace}
\newcommand{\PPINFERFRAME}{\ensuremath{\pi_{\mathsf{FINFER}}}\xspace}

\usepackage{algorithm}
\usepackage{algorithmic}
\newenvironment{myprotocol}[1][htb]{%
    \floatname{algorithm}{Protocol}
   \begin{algorithm}[#1]%
   \footnotesize{}
  }{\end{algorithm}}

\usepackage{amsthm}
\theoremstyle{definition}

\begin{document}

\title{Privacy-Preserving Video Classification with Convolutional Neural Networks}

\author{Sikha Pentyala, Rafael Dowsley and Martine De Cock
\thanks{Sikha Pentyala is with the School of Engineering and Technology, University of Washington, Tacoma, WA, USA. Email: sikha@uw.edu}
\thanks{Rafael Dowsley is with the Faculty of Information Technology, Monash University, Clayton, Australia. Email: rafael.dowsley@monash.edu}
\thanks{Martine De Cock is with the School of Engineering and Technology, University of Washington, Tacoma, WA, USA and Ghent University, Ghent, Belgium. Email: mdecock@uw.edu}
}

\maketitle

\begin{abstract}
 Many video classification applications require access to personal data, thereby posing an invasive security risk to the users' privacy. We propose a privacy-preserving implementation of single-frame method based video classification with convolutional neural networks that allows a party to infer a label from a video without necessitating the video owner to disclose their video to other entities in an unencrypted manner. 
  Similarly, our approach removes the requirement of the classifier owner from revealing their model parameters to outside entities in plaintext. To this end, we combine existing Secure Multi-Party Computation (MPC) protocols for private image classification with our novel MPC protocols for oblivious single-frame selection and secure label aggregation across frames. The result is an end-to-end privacy-preserving video classification pipeline. We evaluate our proposed solution in an application for private human emotion recognition. Our results across  
  a variety of security settings, spanning honest and dishonest majority configurations of the computing parties, and
  for both passive and active adversaries, 
  demonstrate that videos can be classified with state-of-the-art accuracy, and without leaking sensitive user information.  
\end{abstract}

\IEEEpeerreviewmaketitle

%
%

\section{Introduction}\label{intro}

%

Deep learning based video classification is extensively used in a growing variety of applications, such as facial recognition, activity recognition, gesture analysis, behavioral analysis, eye gaze estimation, 
and emotion recognition in empathy-based AI systems \cite{ali2020spatio, Liu_2019_CVPR, LI_2020_WACV,10.1145/3394171.3416298,9137927,9024044,inproceedingsstudentemotion,10.1145/3340555.3353718,Wu_2019_CVPR}.  
%
%
%
 Many existing and envisioned applications of video classification rely on personal data, rendering these applications invasive of privacy. 
 This applies among other tasks to video surveillance and home monitoring systems.
 Similarly, empathy-based AI systems expose personal emotions, which are most private to a user, to the service provider. Video classification systems deployed in commercial applications commonly require user videos to be shared with the service provider or sent to the cloud. These videos may remain 
publicly available on the Internet. Users have no control over the deletion of the videos, and the data may be available for scraping, as done for instance by Clearview AI \cite{ClearViewAI2020}. 
The need to protect the privacy of individuals is widely acknowledged \cite{NPRS2016}.
Concerns regarding privacy of user data are giving rise to new laws and regulations such as the European GDPR and the California Consumer Privacy Act (CCPA), as well as a perceived tension between the desire to protect data privacy on one hand, and to promote an economy based on free-flowing data on the other hand \cite{kalman2018}. The E.U.~is for instance considering a three-to-five-year moratorium on face recognition in public places, given its significant potential for misuse \cite{Chee:2020}.



A seemingly straightforward technique to keep user videos private is to deploy the deep learning models of the service providers at the user-end instead of transferring user data 
to the cloud. This is not a viable solution for several reasons. First, owners of proprietary models are concerned about shielding their model, especially when it constitutes a competitive advantage. Second, in security applications such as facial recognition, or deepfake detection, revealing model details helps adversaries develop evasion strategies. Furthermore, powerful deep learning models that memorize their training examples are well known; one would not want to expose those by revealing the model. Finally, deployment of large deep learning models at the user end may be technically difficult or impossible due to limited computational resources.
For these reasons, ML tasks such as video classification are commonly outsourced to a set of efficient cloud servers in a Machine-Learning-as-a-Service (MLaaS) architecture.
Protecting the privacy of both the users' and the service provider's data while performing outsourced ML computations is an important challenge.


%
Privacy-preserving machine learning (PPML) has been hailed, even by politicians \cite{EvidenceBasedPolicy2017, Wyden2017}, as a potential solution when handling sensitive information. Substantial technological progress has been made during the last decade in the area of Secure Multi-Party Computation (MPC) \cite{CDN2015}, an umbrella term for cryptographic approaches that allow two or more parties to jointly compute a specified output from their private information in a distributed fashion, without revealing the private information to each other. Initial applications of MPC based privacy-preserving inference with deep learning models have been proposed for image  \cite{agrawal2019quotient,dalskov2019secure,juvekar2018gazelle,kumar2020cryptflow,mishra2020delphi,riazi2018chameleon,riazi2019xonn,rouhani2018deepsecure} and audio classification \cite{bittner2020private}.
We build on this existing work to create the first end-to-end MPC protocol for private 
video classification. In our solution, videos are classified according to the well-known \textit{single-frame} method, i.e.~by aggregating predictions across single frames/images.
Our main novel contributions are:
\begin{itemize}[leftmargin=*,noitemsep,topsep=0pt]
\item A protocol for selecting frames in an oblivious manner.
\item A protocol for secure frame label aggregation.
\item An evaluation of our secure video classification pipeline in an application for human emotion detection from video on the RAVDESS dataset, demonstrating that MPC based video classification is feasible today, with state-of-the-art classification accuracies, and without leaking sensitive user information.
\end{itemize}

\begin{figure}
    \centering
    \includegraphics[width=8.5cm]{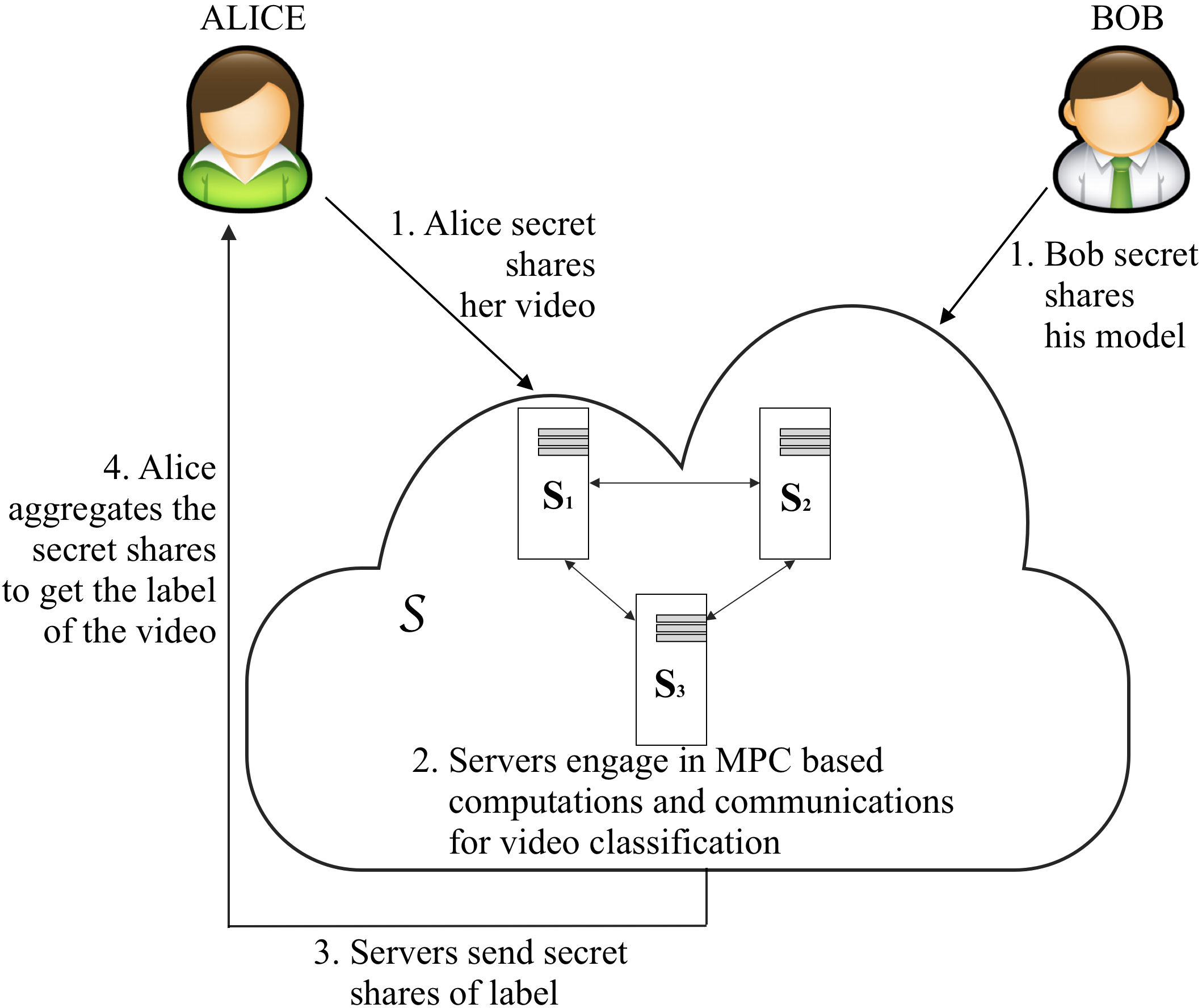}
    \caption{Privacy-preserving video classification as an outsourced computation problem, illustrated for the 3-party computation setting (3PC) with 3 servers $S_1$, $S_2$, and $S_3$}
    \label{fig:obemotion}
\end{figure}

Fig.~\ref{fig:obemotion} illustrates the flow of our proposed solution at a high level. The video of end user \textit{Alice} should be classified with \textit{Bob}'s model in such a way that no one other than Alice sees the video, and no one other than Bob sees the model parameters. Below we refer to both Alice's video and Bob's model parameters as ``data''.
In Step 1 of Fig.~\ref{fig:obemotion}, Alice and Bob each send secret shares of their data to a set $S$ of untrusted servers (``parties''). While the secret shared data can be trivially revealed by combining all shares, nothing about the data is revealed to any subset of the servers that can be corrupted by the adversary. This means, in particular, that none of the servers by themselves learns anything about the actual values of the data. Next, in Step 2, the servers execute MPC protocols for oblivious frame selection, image classification, and frame label aggregation. Throughout this process, none of the servers learns the values of the data nor the assigned label, as all computations are done over secret shares. Finally, in Step 3, the servers can reveal their shares of the computed class label to Alice, who combines them in Step 4 to learn the output of the video classification.

Steps 1 and 3-4 are trivial as they follow directly from the choice of the underlying MPC scheme (see Sec.~\ref{prelim}). The focus of this paper is on Step 2, in which the servers (parties) execute protocols to perform computations over the secret shared data (see Sec.~\ref{methods}). MPC is concerned with the protocol execution coming under attack by an adversary which may corrupt parties to learn private information or cause the result of the computation to be incorrect. MPC protocols are designed to prevent such attacks being successful.
There exist a variety of MPC schemes, designed for different numbers of parties and offering various levels of security that correspond to different threat models, and coming with different computational costs. Regarding threat models, we consider settings with \textit{semi-honest} as well as with \textit{malicious} adversaries. While parties corrupted by semi-honest adversaries follow the protocol instructions correctly but try to obtain additional information, parties corrupted by malicious adversaries can deviate from the protocol instructions. Regarding the \textit{number} of parties (servers), some of the most efficient MPC schemes have been developed for 3 parties, out of which at most one is corrupted. We evaluate the runtime of our protocols in this honest-majority 3-party computing setting (3PC), which is growing in popularity in the PPML literature, e.g.~\cite{dalskov2019secure,kumar2020cryptflow,riazi2018chameleon,wagh2019securenn,patra2020blaze}, and we demonstrate how even better runtimes can be obtained with a recently proposed MPC scheme for 4PC with one corruption \cite{cryptoeprint:2020:1330}.
Our protocols are generic and can be used in a 2PC, dishonest-majority setting as well, i.e.~where each party can only trust itself. Note that in the 2PC setting, the computation can be performed directly by Alice and Bob if they are not very limited in terms of computational resources.  As known from the literature, and apparent from our results, the higher level of security offered by the 2PC setting
comes with a substantial increase in runtime.

After discussing related work in Sec.~\ref{rw} and recalling preliminaries about MPC in Sec.~\ref{prelim}, we present our protocols for privacy-preserving video classification in Sec.~\ref{methods}. 
The MPC protocols we present in Sec.~\ref{methods} enable the servers to perform all these computations without accessing the video $\mathcal{V}$ or the convolutional neural network (ConvNet) model $\mathcal{M}$ in plaintext. In Sec.~\ref{RESULTS} we present an experimental evaluation of our method when applied to emotion recognition from videos of the RAVDESS dataset. Our ConvNet based secure video classification approach achieves accuracies at par with those in the literature for this dataset, while not requiring leakage of sensitive information. Our prototype classifies videos that are 3-5 sec in length in under 14 sec on Azure F32 machines, demonstrating that private video classification based on MPC is feasible today.

\section{Related work}\label{rw}
\textbf{Privacy-preserving video classification.}\label{rwsvc}
Given the invasive nature of video classification applications, it is not surprising that efforts have been made to protect the privacy of individuals.
Non-cryptography based techniques such as anonymizing faces in videos \cite{ren2018learning}, pixel randomization to hide the user's identity \cite{10.1145/3341105.3373942}, compressing video frames to achieve visual shielding effect \cite{liu2020privacy}, lowering resolution of videos \cite{ryoo2016privacypreserving}, using autoencoders to maintain privacy of the user's data \cite{d2020autoencoder}, and changes in ways the videos are captured \cite{wang2019privacypreserving} do not provide any formal privacy guarantees and affect the accuracy of the inference made. Solutions based on \textit{Differential Privacy (DP)} \cite{wang2019privacypreservingdp} introduce noise, or replace the original data at the user end by newly generated data, to limit the amount of information leaked, at the cost of lowering accuracy. The recently proposed ``Visor'' system requires secure hardware (trusted execution environments) for privacy-preserving video analytics \cite{255282}.

In contrast to the approaches above, in this paper we pursue the goal of having \textit{no leakage} of information during the inference phase, \textit{without} requiring special secure hardware. To the best of our knowledge, our approach is the first in the open literature to achieve this goal for private video classification. To this end, we leverage prior work on cryptography based private image classification, as described below, and augment it with novel cryptographic protocols for private video frame selection and label aggregation across frames.  

\textbf{Cryptography based image classification.}\label{rwscnn}
There are 2 main approaches within cryptography that enable computations over encrypted data, namely \textit{Homomorphic Encryption} (HE) and \textit{Secure Multiparty Computation} (MPC).
Both have been applied to secure inference with trained neural networks, including for image classification with ConvNets
\cite{
byali2020flash,
gilad2016cryptonets,
koti2020swift,kumar2020cryptflow,
patra2020blaze,rachuri2019trident,riazi2019xonn,riazi2018chameleon,
wagh2019securenn,wagh2020falcon}. 
Neither have been applied to video classification before.
While HE has a lower communication burden than MPC, it has much higher computational costs, making HE less appealing at present for use in applications where response time matters. E.g., in state-of-the-art work on private image classification with HE, Chillotti et al.~(\cite{cryptoeprint:2021:091}) report a classification time of $\sim 9$ sec for a $28 \times 28$ MNIST image on 96vCPU AWS instances with a neural network smaller in size (number of parameters) than the one we use in this paper. As demonstrated in Sec.~\ref{RESULTS}, the MPC based techniques for image classification based on Dalskov et al.~(\cite{dalskov2019secure}) that we use, can label images (video frames) an order of magnitude faster, even when run on less powerful 32vCPU Azure instances ($\sim 0.5$ sec for passive 3PC; $\sim 1$ sec for active 4PC). We acknowledge that this superior performance stems from the flexibility of MPC to accommodate honest-majority 3PC/4PC scenarios. HE based private image classification is by design limited to the dishonest-majority 2PC setting, in which our MPC approach is too slow for video classification in (near) real-time as well.    

\textbf{Emotion recognition.}\label{rwemot}
A wide variety of applications have prompted research in emotion recognition, using various modalities and features 
\cite{bhattacharya2019step,Jia_2019_CVPR,jiao2019real,
Wei_2020_CVPR}, including videos 
\cite{
zhao2020endtoend,hu2019video,
mittal2019m3er,Mittal_2020_CVPR,deng2019mimamo}. 
Emotion recognition from videos in the RAVDESS benchmark dataset, as we do in the use case in Sec.~\ref{RESULTS}, has been studied by other authors in-the-clear, i.e.~without regards for privacy protection, using a variety of deep learning architectures, with reported accuracies in the 57\%-82\% range, depending on the number of emotion classes included in the study (6 to 8) \cite{bagheri2019novel,mansouri2020synch,bursic2020improving,9051332}.
The ConvNet model that we trained for our experimental results in Sec.~\ref{RESULTS} is at par with these state-of-the-art accuracies.
Jaiswal and Provost \cite{jaiswal2019privacy} have studied privacy metrics and leakages when inferring emotions from data. To the best of our knowledge, there is no existing work on privacy-preserving emotion detection from videos using MPC, as we do in Sec.~\ref{RESULTS}.

%
%

\section{Preliminaries}\label{prelim}
Protocols for Secure Multi-Party Computation (MPC) enable a set of parties to jointly compute the output of a function over the private inputs of each party, without requiring any of the parties to disclose their own private inputs. MPC is concerned with the protocol execution coming under attack by an adversary $\adv$ which may corrupt one or more of the parties to learn private information or cause the result of the computation to be incorrect. MPC protocols are designed to prevent such attacks being successful, and can be mathematically proven to guarantee privacy and correctness.
We follow the standard definition of the Universal Composability (UC) framework \cite{canetti2000security}, in which the security of protocols is analyzed by comparing a real world with an ideal world. For details, see Evans et al.~(\cite{evans2018pragmatic}).


An adversary $\adv$ can corrupt any number of parties. In a \textit{dishonest-majority} setting at least half of the parties are corrupt, while in an \textit{honest-majority} setting, more than half of the parties are honest (not corrupted). Furthermore, $\adv$ can have different levels of adversarial power. In the \textit{semi-honest} model, even corrupted parties follow the instructions of the protocol, but the adversary attempts to learn private information from the internal state of the corrupted parties and the messages that they receive. 
MPC protocols that are secure against semi-honest or \textit{``passive''} adversaries prevent such leakage of information. In the \textit{malicious} adversarial model, the corrupted parties can arbitrarily deviate from the protocol specification. Providing security in the presence of malicious or \textit{``active''} adversaries, i.e.~ensuring that no such adversarial attack can succeed, comes at a higher computational cost than in the passive case. 

The protocols in Sec.~\ref{methods} are sufficiently generic to be used in dishonest-majority as well as honest-majority settings, with passive or active adversaries. This is achieved by changing the underlying MPC scheme to align with the desired security setting. Table \ref{tab:mpcschemes} contains an overview of the MPC schemes used in Sec.~\ref{RESULTS}. In these MPC schemes, all computations are done on integers modulo $q$, i.e.,~in a ring $\mathbb{Z}_q =\{0,1,\ldots,q-1\}$, with $q$ a power of 2. 
The pixel values in \textit{Alice}'s video and the model parameters in \textit{Bob}'s classifier are natively real numbers represented in a floating point format. As is common in MPC, they are converted to integers using a fixed-point representation \cite{FC:CatSax10}. When working with fixed-point representations with $a$ fractional bits, every multiplication generates an extra $a$ bits of unwanted fractional representation. To securely ``chop off'' the extra fractional bits generated by multiplication, we use 
the deterministic truncation protocol 
by Dalskov et al.~(\cite{dalskov2019secure,cryptoeprint:2020:1330}) for computations over $\mathbb{Z}_{2^k}$. 
Below we give a high level description of the 3PC schemes from Table \ref{tab:mpcschemes}. For more details and a description of the other MPC schemes, we refer to the papers in Table \ref{tab:mpcschemes}.

\begin{table}
    \centering
    \begin{tabular}{cc|ll}
    & &  \footnotesize MPC scheme  & \footnotesize Reference\\
    \hline
    \multirow{2}{*}{passive} 
                           & \footnotesize 2PC & \footnotesize OTSemi2k &         \begin{tabular}[c]{@{}l@{}}\footnotesize    semi-honest adaptation of\\                     \footnotesize \cite{cryptoeprint:2018:482}
                           \end{tabular}\\
                             &  \footnotesize 3PC & \footnotesize Replicated2k & \footnotesize \cite{araki2016high} \\
    \hline
    \multirow{3}{*}{active}  
                            & \footnotesize 2PC & \footnotesize SPDZ2k &   
    \begin{tabular}[c]{@{}l@{}}\footnotesize \cite{cryptoeprint:2018:482},\\ \footnotesize \cite{damgaard2019new}\end{tabular}\\
                            & \footnotesize 3PC & 
                                 \begin{tabular}[c]{@{}l@{}}\footnotesize SPDZ-wise\\\footnotesize Replicated2k \end{tabular} & 
                                 \footnotesize \cite{cryptoeprint:2020:1330}\\
                            & \footnotesize 4PC & \footnotesize Rep4-2k & \footnotesize \cite{cryptoeprint:2020:1330}\\
    \end{tabular}
    \caption{MPC schemes used in the experimental evaluation for 2PC (dishonest majority) and 3PC/4PC (honest majority)} 
    \label{tab:mpcschemes}
\end{table}

\textbf{Replicated sharing (3PC).}
After \textit{Alice} and \textit{Bob} have converted all their data to integers modulo $q$, they send secret shares of these integers to the servers in $S$ (see Fig.~\ref{fig:obemotion}).
In a replicated secret sharing scheme with 3 servers (3PC), a value $x$ in $\mathbb{Z}_q$ is secret shared among servers (parties) $S_1, S_2,$ and $S_3$ by picking uniformly random shares $x_1, x_2, x_3 \in \mathbb{Z}_q$ such that 
$x_1 + x_2 +x_3 =  x \mod{q}$,
and distributing $(x_1,x_2)$ to $S_1$, $(x_2,x_3)$ to $S_2$, and $(x_3,x_1)$ to $S_3$. Note that no single server can obtain any information about $x$ given its shares. We use $[\![x]\!]$ as a shorthand for a secret sharing of $x$. The servers subsequently classify Alice's video with Bob's model by computing over the secret sharings.

\textbf{Passive security (3PC).} 
The 3 servers can perform the following operations through carrying out local computations on their own shares: addition of a constant, addition of secret shared values, and multiplication by a constant. For multiplying secret shared values $[\![x]\!]$ and $[\![y]\!]$, we have that $x \cdot y=(x_1 + x_2 +x_3)(y_1 + y_2 +y_3)$, and so $S_1$ computes $z_1=x_1 \cdot y_1+x_1 \cdot y_2+x_2 \cdot y_1$, $S_2$ computes $z_2=x_2 \cdot y_2+x_2 \cdot y_3+x_3 \cdot y_2$ and $S_3$ computes $z_3=x_3 \cdot y_3+x_3 \cdot y_1+x_1 \cdot y_3$. Next, the servers obtain an additive secret sharing of $0$ by picking uniformly random $u_1,u_2,u_3$ such that $u_1 + u_2 +u_3 = 0$, which can be locally done with computational security by using pseudorandom functions, and $S_i$ locally computes $v_i=z_i+u_i$. Finally, $S_1$ sends $v_1$ to $S_3$, $S_2$ sends $v_2$ to $S_1$, and $S_3$ sends $v_3$ to $S_2$, enabling the servers $S_1, S_2$ and $S_3$ to get the replicated secret shares $(v_1,v_2)$, $(v_2,v_3)$, and $(v_3,v_1)$, respectively, of the value $v=x \cdot y$. This protocol only requires each server to send a single ring element to one other server, and no expensive public-key encryption operations (such as homomorphic encryption or oblivious transfer)
are required. This MPC scheme was introduced by Araki et al. (\cite{araki2016high}).

\textbf{Active security (3PC).} 
In the case of malicious adversaries, the servers are prevented from deviating from the protocol and gain knowledge from another party through the use of information-theoretic message authentication codes (MACs). 
For every secret share, an authentication message is also sent to authenticate that each share has not been tampered in each communication between parties. In addition to computations over secret shares of the data, the servers also need to update the MACs appropriately, and the operations are more involved than in the passive security setting. For each multiplication of secret shared values, the total amount of communication between the parties is greater than in the passive case. 
%
We use the MPC scheme \textit{SPDZ-wiseReplicated2k} recently proposed by Dalskov et al.~(\cite{cryptoeprint:2020:1330}), with the option with preprocessing for generation of the multiplication triples that is available in MP-SPDZ \cite{cryptoeprint:2020:521}.

\paragraph{MPC primitives.}
The MPC schemes listed above provide a mechanism for the servers to perform cryptographic primitives through the use of secret shares, namely addition of a constant, multiplication by a constant, and addition and multiplication of secret shared values. Building on these cryptographic primitives, MPC protocols for other operations have been developed in the literature. 
We use:

\begin{itemize}[leftmargin=*,noitemsep,topsep=0pt]
    \item Secure matrix multiplication \pmmul : at the start of this protocol, the parties have secret sharings $[\![A]\!]$ and $[\![B]\!]$ of matrices $A$ and $B$; at the end of the protocol, the parties have a secret sharing $[\![C]\!]$ of the product of the matrices, $C=A \times B$. \pmmul is a direct extension of the secure multiplication protocol for two integers explained above, which we will denote as \pmul in the remainder. 
    
    
    
    
    \item Secure comparison protocol \plt \cite{catrina2010improved}: at the start of this protocol, the parties have secret sharings $[\![x]\!]$ and $[\![y]\!]$ of integers $x$ and $y$; at the end of the protocol they have a secret sharing of $1$ if $x < y$, and a secret sharing of $0$ otherwise.
    
    \item Secure argmax \pargmax : this protocol accepts secret sharings of a vector of integers and returns a secret sharing of the index at which the vector has the maximum value. \pargmax is straightforwardly constructed using the above mentioned secure comparison protocol.
    
    \item Secure RELU \prelu \cite{dalskov2019secure}: at the start of this protocol, the parties have a secret sharing of $z$; at the end of the protocol, the parties have a secret sharing of the value $\max(0,z)$. $\prelu$ is constructed from \plt, followed by a secure multiplication to either keep the original value $z$ or replace it by zero in an oblivious way.
    
    \item Secure division \pdiv : for secure division, the parties use an iterative algorithm that is well known in the MPC literature \cite{catrina2010secure}.
\end{itemize}



%
%
\section{Methodology}\label{methods}


The servers perform video classification based on the single-frame method, i.e.~by (1) selecting frames from the video $\mathcal{V}$ (Sec.~\ref{SEC:SELECTFRAMES}); 
(2) labeling the selected frames with a ConvNet model $\mathcal{M}$ (Sec.~\ref{SEC:LABELFRAMES}); and 
(3) aggregating the labels inferred for the selected frames into a final label for the video (Sec.~\ref{SEC:LABELAGGR}). 
The video $\mathcal{V}$ is owned by \textit{Alice} and the model $\mathcal{M}$ is owned by \textit{Bob}. Neither party is willing or able to reveal their video/model to other parties in an unencrypted manner.

\subsection{Oblivious frame selection}\label{SEC:SELECTFRAMES}

\begin{figure}
    \centering
    \includegraphics[width=8.5cm]{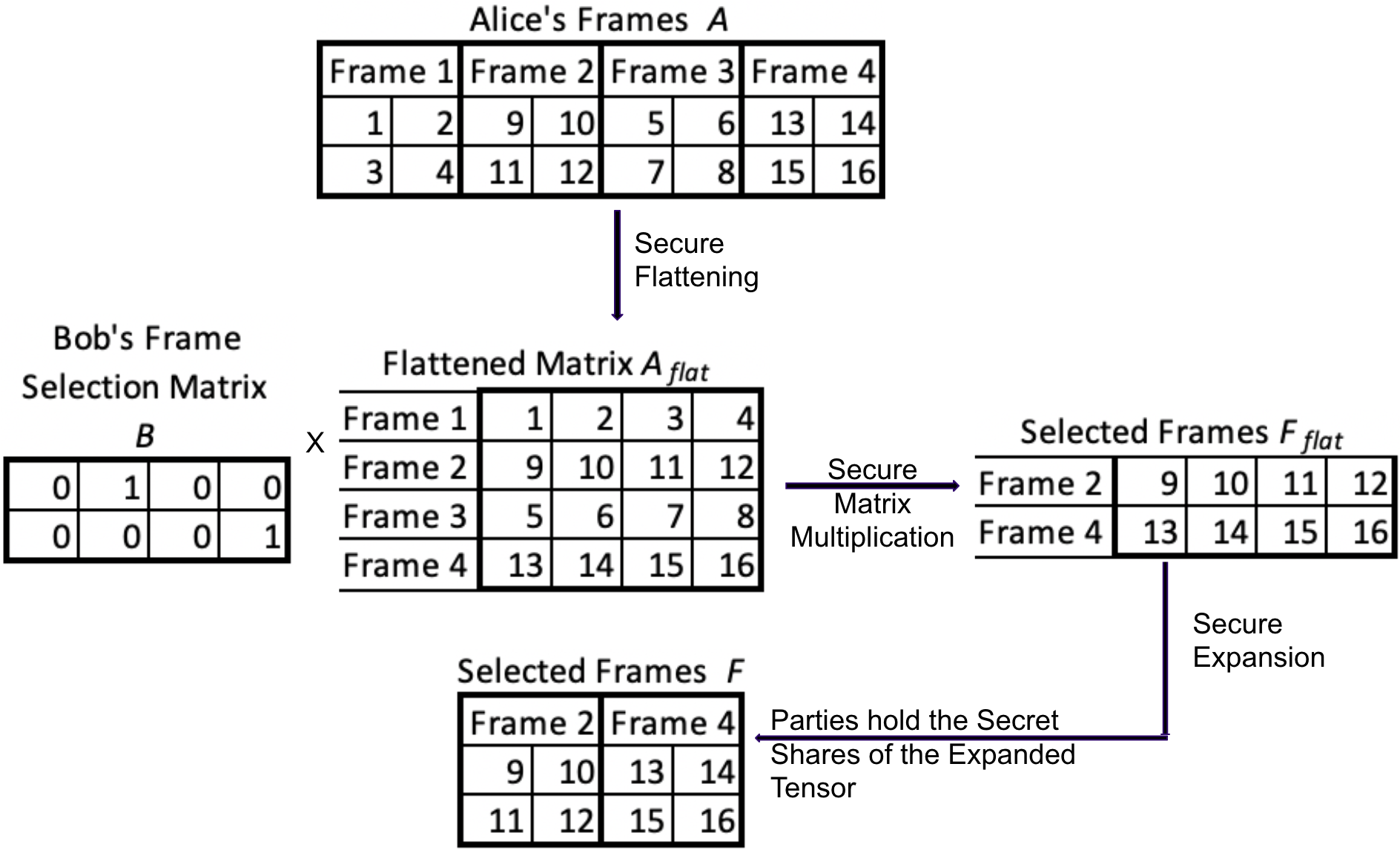}

    \caption{Illustration of oblivious frame selection. The assumption is made that \textit{Alice} has 4 frames in total, each of size $2 \times 2 \times 1$, and \textit{Bob} needs to select 2 frames, namely Frames 2 and 4. \textit{Alice} has a tensor $A$ of size $4 \times 2 \times 2 \times 1$ and \textit{Bob} has a 2D-matrix $B$ of size $2 \times 4$. $A$ is flattened securely to form $A_\mathsf{flat}$ of size $4 \times 4$. A secure matrix multiplication $B \times A_\mathsf{flat}$ is performed  resulting in $F_\mathsf{flat}$, a $2 \times 4$ matrix holding the 2 selected frames. This matrix is then expanded to matrix $F$ of size $2 \times 2 \times 2 \times 1$.}
    \label{fig:frame_selection}
\end{figure}

We assume that \textit{Alice} has prepared her video $\mathcal{V}$ as a 4D array (tensor) $A$ of size $N \times h \times w \times c$ where $N$ is the number of frames, 
$h$ is the height and $w$ is the width of the frame, and $c$ represents the number of color channels of the frame. 
As explained in Sec.~\ref{prelim}, Alice has converted the pixel values into integers using a fixed-point representation.
The values of the dimensions $N, h, w, c$ are known to \textit{Bob} and the set of servers \textit{S}. 
All other properties of the video are kept private, including the video length, the frames per second (fps), and video capture details such as the type of camera used.
Moreover, \textit{Bob} and the servers \textit{S} do not learn the values of the pixels, i.e.~the actual contents of the frames remain hidden from \textit{Bob} and \textit{S} (and anyone else, for that matter). For an illustration of \textit{Alice}'s input, we refer to the top of Fig.~\ref{fig:frame_selection}, where $N=4$, $h=2$, $w=2$, and $c=1$.

\textit{Bob} samples a fixed number of frames from \textit{Alice}'s video, without revealing to \textit{Alice} the frames he is selecting, as such knowledge might allow \textit{Alice} to insert malicious frames in the video in the exact positions that \textit{Bob} is sampling. We assume that \textit{Bob} has a vector $b$ of length $n$, with the indices of the $n$ frames he wishes to select. These indices can for instance be 
$1, 1+d, 1 +2d, \ldots$ for a fixed window size $d$ that is known to \textit{Bob}. In the example in Fig.~\ref{fig:frame_selection}, $n=2$, both 2nd and 4th frames are selected.

The idea behind protocol $\pfselect$ for oblivious frame selection, as illustrated in Fig.~\ref{fig:frame_selection}, is to flatten $A$ into a matrix that contains one row per frame, use a matrix $B$ with one-hot-encodings of the selected frames, multiply $B$ with $A$, and finally expand the product. In more detail: \textit{Bob} converts each entry $i$ of list $b$ (which is an index of a frame to be selected)
into a vector of length $N$ that is a one-hot-encoding of $i$, and inserts it as a row in matrix $B$ of size $n \times N$. \textit{Alice} and \textit{Bob} then send secret shares of their respective inputs $A$ and $B$ to the servers \textit{S}, using a secret sharing scheme as mentioned in Sec.~\ref{prelim}. None of the servers can reconstruct the values of $A$ or $B$ by using only its own secret shares. 

Next the parties in \textit{S} jointly execute protocol $\pfselect$ for oblivious frame selection (see Protocol \ref{prot:fs}). On line 1, the parties reorganize the shares of tensor $A$ of size $N \times h \times w \times c$ into a flattened matrix $A_\mathsf{flat}$ of size $N \times (h \cdot w \cdot c)$. On line 2, the parties multiply $[\![B]\!]$ and $[\![A_\mathsf{flat}]\!]$, using protocol $\pmmul$ for secure matrix multiplication, to select the desired rows from $A_\mathsf{flat}$. On line 3, these selected rows are expanded again into a secret-shared tensor $F$ of size $n \times h \times w \times c$ that holds the selected frames. $F[1], F[2], \ldots, F[n]$ are used in the remainder to denote the individual frames contained in $F$. Throughout this process, the servers do not learn the pixel values from $A$, nor which frames were selected.

\begin{myprotocol}
   \caption{Protocol $\pfselect$ for oblivious frame selection}
   \label{prot:fs}
    \textbf{Input:} A secret shared 4D-array $[\![A]\!]$ of size $N \times h \times w \times c$ with the frames of a video; a secret shared frame selection matrix $[\![B]\!]$ of size $n \times N$. The values $N$, $h$, $w$, $c$, $n$ are known to all parties.
    
    \textbf{Output:} A secret shared 4D-array $F$ of size $n \times h \times w \times c$ holding the selected frames
\begin{algorithmic}[1]
   \STATE $[\![A_\mathsf{flat}]\!]$ $\leftarrow$\\ $\mathsf{RESHAPE}([\![A]\!], N \times h \times w \times c, N \times (h \times w \times c))$
   \STATE $[\![F_\mathsf{flat}]\!]$ $\leftarrow$ $\pmmul([\![B]\!],[\![A_\mathsf{flat}]\!])$
   \STATE $[\![F]\!]$ $\leftarrow$ $\mathsf{RESHAPE}([\![F_\mathsf{flat}]\!], n \times (h \times w \times c), n \times h \times w \times c)$
   \STATE \textbf{return} $[\![F]\!]$
\end{algorithmic}
\end{myprotocol}

\subsection{Private frame classification}\label{SEC:LABELFRAMES}
\label{SEC:ARCH}

We assume that \textit{Bob} has trained an ``MPC-friendly'' 2D-ConvNet $\mathcal{M}$ for classifying individual video frames (images), and that \textit{Bob} secret shares the values of the model parameters with the servers \textit{S}, who already have secret shares of the selected frames from \textit{Alice}'s video after running Protocol $\pfselect$.
By ``MPC-friendly'' we mean that the operations to be performed when doing inference with the trained ConvNet are chosen purposefully among operations for which efficient MPC protocols exist or can be constructed. Recall that a standard ConvNet contains one or more blocks that each have a convolution layer, followed by an activation layer, typically with RELU, and an optional pooling layer. These blocks are then followed by fully connected layers which commonly have RELU as activation function, except for the last fully connected layer which typically has a Softmax activation for multi-class classification. The operations needed for all layers, except for the output layer, boil down to comparisons, multiplicatons, and summations. All of these cryptographic primitives can be efficiently performed with state-of-the-art MPC schemes, as explained in Sec.~\ref{prelim}. Efficient protocols for convolutional, RELU activation, average pooling layers, and dense layers are known in the MPC literature \cite{dalskov2019secure}. We do not repeat them in this paper for conciseness. All these operations are performed by the servers \textit{S} using the secret shares of \textit{Bob}'s model parameters and of the selected frame from \textit{Alice}'s video, as obtained using $\pfselect$.



As previously mentioned, Softmax is generally used as the activation function in the last layer of ConvNets that are trained to perform classification. 
Softmax normalizes the logits passed into it from the previous layer to a probability distribution over the class labels. Softmax is an expensive operation to implement using MPC protocols, as this involves division and exponentiation. Previously proposed workarounds include disclosing the logits and computing Softmax in an unencrypted manner  \cite{liu2017oblivious}, which leaks information, or replacing Softmax by Argmax \cite{bittner2020private,dalskov2019secure}. The latter works when one is only interested in retrieving the class label with the highest probability, as the Softmax operation does not change the ordering among the logits. In our context of video classification based on the single-frame method however, the probabilities of all class labels for each frame are required, to allow probabilities across the different frames to be aggregated to define a final label (see Sec.~\ref{SEC:LABELAGGR}).

        
        
        
        
        



We therefore adopt the solution proposed by Mohassel and Zhang \cite{mohassel2017secureml} and replace the Softmax operation by
$$
\scriptsize{
f(u_i) = 
\left\{
\begin{array}{ll}
\displaystyle \frac{\mbox{RELU}(u_i)}{\sum\limits_{j=1}^{C}\mbox{RELU}(u_j)}, & \mbox{if\ } \sum\limits_{j=1}^{C}\mbox{RELU}(u_j) > 0\\
\\
1/C, & \mbox{otherwise}
\end{array}
\right.
}
$$
for $i=1,\ldots, C$, where $(u_1,u_2,\ldots,u_C)$ denote the logits for each of the $C$ class labels, and $(f(u_1),f(u_2),\ldots,f(u_C))$ is the computed probability distribution over the class labels.
Pseudocode for the corresponding MPC protocol is presented in Protocol \ref{prot:approxSM}. At the start of Protocol $\pisoft$, the servers have secret shares of a list of logits, on which they apply the secure RELU protocol in Line 1. Lines 2-5 serve to compute the sum of the RELU values, while on Line 6 the parties run a secure comparison protocol to determine if this sum is greater than 0. If $Sum_\mathsf{relu}$ is greater than 0, then after Line 6, $[\![cn]\!]$ contains a secret sharing of 1; otherwise it contains a secret sharing of 0. Note that if $cn = 1$ then the numerator of the $i^{th}$ probability $f(u_i)$ should be $X_\mathsf{relu}[i]$ while the denominator should be $Sum_\mathsf{relu}$. Likewise, if $cn = 0$ then the numerator should be $1$ and the denominator $C$. As is common in MPC protocols, we use multiplication instead of control flow logic for such conditional assignments. To this end, a conditional based branch operation as ``$\textbf{if  } p \textbf{  then  } q \leftarrow r \textbf{  else  } q \leftarrow s$'' is rephrased as 
``$q  \gets  p \cdot r  +  (1-p)  \cdot  s$''.
In this way, the number and the kind of operations executed by the parties does not depend on the actual values of the inputs, so it does not leak information that could be exploited by side-channel attacks.
Such conditional assignments occur in Line 7 and 10 of Protocol $\pisoft$, to assign the correct value of the numerator and the denominator.

\begin{myprotocol}
\caption{Protocol $\pisoft$ for approximate Softmax}
   \label{prot:approxSM}
    \textbf{Input:} A secret shared list $[\![logits]\!]$ of logits of size $C$, where $C$ is total number of class labels
    
    \textbf{Output:} A secret shared list $[\![SM_\mathsf{approx}]\!]$ of size $C$ of probabilities for the class labels
\begin{algorithmic}[1]
     \STATE $[\![X_\mathsf{relu}]\!]$ $\leftarrow$ \prelu($[\![logits]\!]$)\\
        
     \STATE $[\![Sum_\mathsf{relu}]\!]$ $\leftarrow$ $0$\\
  
        \FOR{$j=1$ {\bfseries to} $C$} {
        \STATE $[\![Sum_\mathsf{relu}]\!]$ $\leftarrow$ $[\![Sum_\mathsf{relu}]\!]$ + $[\![X_\mathsf{relu}[i]]\!]$
        }
        \ENDFOR
        
        \STATE $[\![cn]\!]$ $\leftarrow$ \plt($0$,$[\![Sum_\mathsf{relu}]\!]$)
        
        \STATE $[\![denom]\!]$ $\leftarrow$ \pmul($[\![cn]\!]$, $[\![Sum_\mathsf{relu}]\!]$)  +       \pmul(($1 - [\![cn]\!]$), $C$)
        
        \STATE $[\![denom\_inv]\!]$ $\leftarrow$ \pdiv($1, [\![denom]\!])$

        \FOR{$i=1$ {\bfseries to} $C$} {
        \STATE $[\![numer]\!]$ $\leftarrow$ \pmul($[\![cn]\!]$, $[\![X_\mathsf{relu}[i]]\!]$) + ($1 - [\![cn]\!]$) \\
        $[\![SM_\mathsf{approx}[i]]\!]$ $\leftarrow$ \pmul($[\![numer]\!]$, $[\![denom\_inv]\!]$) \\
        }
        \ENDFOR
        \STATE \textbf{return} $[\![SM_\mathsf{approx}]\!]$
\end{algorithmic}
\end{myprotocol}

A protocol \pifi for performing secure inference with \textit{Bob}'s model $\mathcal{M}$ (which is secret shared among the servers) over a secret shared frame $f$ from \textit{Alice}'s video can be straightforwardly obtained by: (1) using the cryptographic primitives defined in Sec.~\ref{prelim} to securely compute all layers except the output layer; (2) using Protocol $\pisoft$ to compute the approximation of the Softmax for the last layer. The execution of this protocol results in the servers obtaining secret shares of the inferred probability distribution over the class labels for frame $f$.

\subsection{Secure label aggregation}
\label{SEC:LABELAGGR}





\begin{figure}
    \centering
    \includegraphics[width=0.8\columnwidth]{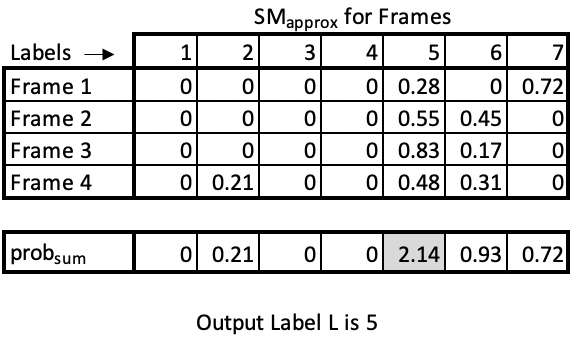}
    \caption{Illustration of label aggregation. Let us assume that $n$ = 4 frames were selected for secure inference, and that there are $C$ = 7 classes. $SM_\mathsf{approx}$ holds the inferred probability distribution over the class labels for each frame. Class label $5$ is selected as the final label because it has the highest sum of probabilities across all classified frames.}
    \label{fig:securelabelagg}
\end{figure}

As illustrated in Fig.~\ref{fig:securelabelagg}, we aggregate the predictions across the single frames by selecting the class label with the highest sum of inferred probabilities across the frames.
We implement this securely as Protocol 
\ref{prot:vcsecure}. To classify a video $\mathcal{V}$, the servers: (1) obliviously select the desired frames as shown in Line 2; (2) securely infer the probability distribution $SM_\mathsf{approx}$ of all classes labels generated by the model $\mathcal{M}$ on a specific selected frame, as shown in Line 4; (3) add these probabilities, index-wise, to the sum of the probabilities corresponding to each class that is obtained throughout the selected frames (Line 5-6); (4) securely find the index $L$ with maximum value in the aggregated list (Line 8). $L$ represents the class label for the video. At the end of Protocol \ref{prot:vcsecure}, the servers hold a secret sharing $[\![L]\!]$ of the video label. Each of the servers sends its secret shares to \textit{Alice}, who uses them to construct the class label $L$ for the video.

\begin{myprotocol}
\caption{Protocol \pvc for classifying a video securely based on the single-frame method}
   \label{prot:vcsecure}
    \textbf{Input:} A video $\mathcal{V}$ secret shared as a 4D-array $[\![A]\!]$, a frame selection matrix secret shared as $[\![B]\!]$, the parameters of the ConvNet model $\mathcal{M}$ secret shared as $[\![M]\!]$
    
    \textbf{Output:} A secret share $[\![L]\!]$ of the video label
\begin{algorithmic}[1]
     \STATE Let $[\![prob_\mathsf{sum}]\!]$ be a list of length $C$ that is initialized with zeros in all indices.
     \STATE $[\![F]\!]$ $\leftarrow$  \pfselect($[\![A]\!]$, $[\![B]\!]$)\\
    
    \FORALL{$[\![F[j]]\!]$}{
    \STATE $[\![SM_\mathsf{approx}]\!]$ $\leftarrow$ \pifi($[\![M]\!]$, $[\![F[j]]\!]$)
    \FOR{$i=1$ {\bfseries to} $C$} {
    \STATE $[\![prob_\mathsf{sum}[i]]\!]$ $\leftarrow$ $[\![prob_\mathsf{sum}[i]]\!]$ + $[\![SM_\mathsf{approx}[i]]\!]$
    }
    \ENDFOR 
    }
    \ENDFOR    
    \STATE $[\![L]\!]$ $\leftarrow$ \pargmax($[\![prob_\mathsf{sum}]\!]$)

    \STATE \textbf{return} $[\![L]\!]$
\end{algorithmic}
\end{myprotocol}

%
%
\section{Results}\label{RESULTS}

\subsection{Dataset and model architecture}
\label{RESULT_DS}
We demonstrate the feasibility of our privacy-preserving video classification approach for the task of emotion detection using the 
database \footnote{\url{https://zenodo.org/record/1188976}} \cite{livingstone2018ryerson}. 
We use 1,248 video-only files with speech modality from this dataset, corresponding to 7 different emotions, namely \textit{neutral} (96), \textit{happy} (192), \textit{sad} (192),  \textit{angry} (192), \textit{fearful} (192), \textit{disgust} (192), and \textit{surprised} (192). The videos portray 24 actors who each read two different statements twice, with different emotions, for a total of 52 video files per actor. For all emotions except for neutral, the statements are read with alternating normal and strong intensities; this accounts for the fact that there are less ``neutral'' instances in the dataset than for the other emotion categories. As in  \cite{bursic2020improving}, we leave out the \textit{calm} instances, reducing the original 8 emotion categories from the RADVESS dataset to the 7 categories that are available in the FER2013 dataset \cite{carrier2013fer}, which we use for pre-training.
%
The videos in the RAVDESS dataset have a duration of 3-5 seconds with 30 frames per second, hence the total number of frames per video is in the range of 120-150. 
We split the data into 1,116 videos for training and 132 videos for testing. To this end, we moved all the video recordings of the actors 8, 15 (selected randomly) and an additional randomly selected 28 video recordings to the test set,  while keeping the remaining video recordings in the train set.

We used OpenCV \cite{opencv_library} to read the videos into frames. 
Faces are detected with a confidence greater than 98\% using MTCNN \cite{zhang2016joint}, aligned, cropped, and converted to gray-scale. Each processed frame is resized to $48 \times 48$, reshaped to a 4D-array, and normalized by dividing each pixel value by $255$. 

\label{RESULT_DPBOB}
For \textit{Bob}'s image classification model, we trained a ConvNet with $\sim1.48$ million parameters with an architecture of [(CONV-RELU)-POOL]-[(CONV-RELU)*2-POOL]*2-[FC-RELU]*2-[FC-SOFTMAX]. 
We pre-trained\footnote{With early stopping using a batch size of 256 and Adam optimizer with default parameters in Keras \cite{chollet2015keras}.} the feature layers on the FER 2013 data 
to learn to extract facial features for emotion recognition,
and fine-tuned\footnote{With early-stopping using a batch size of $64$ and SGD optimizer with a learning rate $0.001$, decay as $10^{-6}$, and momentum as $0.9$.} the model on the RAVDESS training data.
%
%
%
%
Our video classifier samples every 15th frame, classifies it with the above ConvNet, and assigns as the final class label the label that has the highest average probability across all frames in the video. The video classification accuracy on the test set is 56\%. For inference with the MPC protocols, after training, we replace the Softmax function on the last layer by the approximate function discussed in Section \ref{SEC:LABELFRAMES}. After this replacement, the accuracy of the video classifier is 56.8\%.
This is in line with state-of-the-art results in the literature on emotion recognition from RAVDESS videos, namely 57.5\% with Synchronous Graph Neural Networks (8 emotions) \cite{mansouri2020synch}; 61\%  with ConvNet-LSTM (8 emotions) \cite{9051332}; 59\% with an RNN (7 emotions) \cite{bursic2020improving}, and 82.4\% with stacked autoencoders (6 emotions) \cite{bagheri2019novel}.

\subsection{Runtime experiments}\label{prelims_settings}
We implemented the protocols from Sec.~\ref{methods} in the MPC framework MP-SPDZ \cite{cryptoeprint:2020:521}, and ran experiments on co-located F32s V2 Azure virtual machines. Each of the parties (servers) ran on separate VM instances (connected with a Gigabit Ethernet network), which means that the results in Table \ref{tab:finalresults} 
cover communication time in addition to computation time. 
A F32s V2 virtual machine contains 32 cores, 64 GiB of memory, and network bandwidth of upto 14 Gb/s. 
For the ring $\mathbb{Z}_{2^k}$, we used value $k=64$. 



\begin{table}
\centering
\begin{tabular}{l|c||r|r}
\multicolumn{2}{c||}{} & Avg. Time (sec) & Avg. Comm (GB)\\
\hline
\multirow{2}{*}{Passive} & 2PC & 511.64  &  669.35\\
                         & 3PC & \textbf{13.12 } & 2.49 \\
\hline
\multirow{3}{*}{Active}  & 2PC & 8423.81 & 7782.74 \\ 
                         & 3PC  & 48.20 & 10.98  \\ 
                         & 4PC  & \textbf{18.40 } & 4.60 \\                                                                         
\end{tabular}
\caption{Average time to privately detect emotion in a RAVDESS video of duration 3-5 seconds. The average time is computed over a set of 10 videos with a number of frames in the 7-10 range, and with n\_threads=32 in MP-SPDZ. Communication is measured per party.}
\label{tab:finalresults}
\end{table}

\label{RESULT_DIS}
Table \ref{tab:finalresults} presents the average time needed to privately classify a video. 
The MPC schemes for 4PC (with one corrupted party) are faster than 3PC (with one corrupted party), which are in turn substantially faster than 2PC. 
Furthermore, as expected, there is a substantial difference in runtime between the semi-honest (passive security) and malicious (active security) settings. These findings are in line with known results from the MPC literature \cite{cryptoeprint:2020:1330,dalskov2019secure}.
%
%
In the fastest setting, namely a 3PC
setting with a semi-honest adversary that can only corrupt one party, videos from the RAVDESS dataset are classified on average in 13.12 sec, which corresponds to approximately 0.5-0.6 sec per frame, demonstrating that privacy-preserving video classification with state-of-the-art accuracy is feasible in practice. While the presented runtime results are still too slow for video classification in real-time, there is a clear path to substantial optimization that would enable deployment of our proposed MPC solution in practical real-time applications. Indeed, MPC schemes are normally divided in two phases: the offline and online phases. The runtime results in Table \ref{tab:finalresults} represent the time needed for both. The offline phase only performs computations that are independent from the specific inputs of the parties to the protocol (\textit{Alice}'s video and \textit{Bob}'s trained model parameters), and therefore can be executed long before the inputs are known. By executing the offline phase of the MPC scheme in advance, it is possible to improve 
the responsiveness of the final solution.


%
%
\section{Conclusion and Future Work}\label{CONC}
We presented the first end-to-end solution for private video classification based on Secure Multi-Party Computation (MPC). To achieve state-of-the-art accuracy while keeping our architecture lean, we used the single-frame method for video classification with a ConvNet. To keep the videos and the model parameters hidden, we proposed novel MPC protocols for oblivious frame selection and secure label aggregation across frames. We used these in combination with existing MPC protocols for secure ConvNet based image classification, 
and evaluated them for the task of emotion recognition from videos in the RAVDESS dataset.

Our work provides a baseline for private video classification based on cryptography. It can be improved and adapted further to align with state-of-the-art techniques in video classification in-the-clear, including the use of machine learning for intelligent frame selection. While our approach considers only spatial information in the videos, the model architecture in Sec.~\ref{SEC:ARCH} can be replaced by different architectures such as CONV3D, efficient temporal modeling in video \cite{lin2019tsm}, single and two stream ConvNets \cite{Karpathy_2014_CVPR,NIPS2014_5353}  to fuse temporal information. 
Many such approaches use popular ImageNet models for which efficient MPC protocols are available in the literature \cite{dalskov2019secure,kumar2020cryptflow}, opening up interesting directions for further research.

\bibliographystyle{plain}
\bibliography{references}

\end{document}